# A 915-1220 TOPS/W Hybrid In-Memory Computing based Image Restoration and Region Proposal Integrated Circuit for Neuromorphic Vision Sensors in 65nm CMOS

Xueyong Zhang[1], Arindam Basu[2]
[1] Nanyang Technological University (NTU), Singapore
[2] NTU, Singapore and City University of Hong Kong

*Abstract*—In this work, we present a hybrid memory bit cell - collocated SRAM and DRAM (CRAM) consisting of 11 transistors for in-memory computing (IMC) based image restoration (IR) and region proposal (RP). A robust RP updated algorithm is proposed to improve the performance. This work demonstrates IMC based global parallel diffusion and column/row-wise projection to achieve a maximal energy efficiency of 1220 TOPS/W for image restoration and 915 TOPS/W when combined with region proposal.

## I. INTRODUCTION

Bio-inspired asynchronous event-based neuromorphic vision sensors (NVS) are introducing a paradigm shift in visual information sensing and processing [1]. The feature of event-driven operation makes it ideal for low-power operation in the Internet-of-Things scenario such as traffic monitoring. However, the inherent noise in the sensor causes redundant wake-up operation and reduces tracking performance [2]. Energy efficient in-memory computing (IMC) based denoise operation allows blank-frame detection to gain 2X energy savings. Further energy savings can be obtained by exploiting spatial redundancy—objects usually occupy a small part ~5% of the frame in traffic monitoring [3]. Hence, region proposal (RP) is required to detect the region of interests (ROIs) in a valid frame along with their bounding box location coordinates, as shown in Fig. 1. For binary images, the conventional connected component labeling (CCL) algorithm [4] can propose ROIs by raster scanning the whole frame, but leads to longer search time and higher computing energy due to von Neumann operation. The promising IMC approach [3] has high energy efficiency, but has limited accuracy due to a simple algorithm constrained by in-memory operations as well as object fragmentation due to smooth surfaces (e.g. car windows) that do not generate events.

In this work, we present a hybrid memory bit cell—collocated SRAM and DRAM (CRAM) consisting of 11 transistors for IMC-based image restoration (IR) and RP. The proposed CRAM supports image storage in SRAM and DRAM modes, denoise and region filling in diffusion mode and RP algorithm in projection mode.

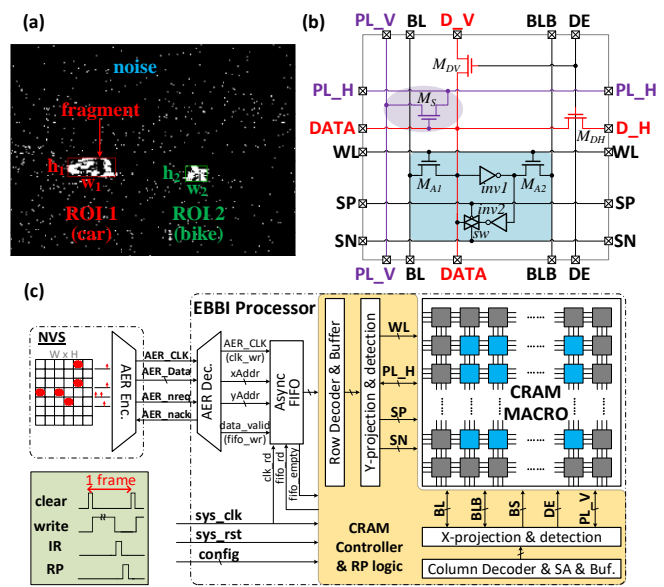

Fig. 1. (a) A typical noisy event-based binary image (EBBI). (b) bit cell of the proposed CRAM. (c) The top-level architectural diagram.

## II. ARCHITECTURE

Fig. 1 shows the proposed CRAM bit-cell and the chip architecture. The versatile CRAM can be divided into three parts: (1) a 6T SRAM (inv1, inv2, MA1, and MA2) with a transmission gate (TG) switch (sw) inserted after the output of inverter inv2 and before the storage node, (2) the two vertical and horizontal NMOS MDV and MDH are employed to connect its 2-D neighbors when in IMC based IR mode, and (3) an NMOS transistor MS works together with access transistor MA1 as a conventional 1T1C DRAM. Moreover, MS also acts as a projection device used for IMC-based global diffusion and column/row-wise projection for RP. The chip includes an address event representation (AER) decoder, a 128x32-bit asynchronous buffer, a 320x240 CRAM macro, and the controller module. It supports four primary operation modes, i.e., clear, write, IR, and RP. The X/Y-projection and detection module finds objects locations along x-axis and y-axis by pulling up the projection lines (PLs) in one direction and sensing PLs from the other direction.

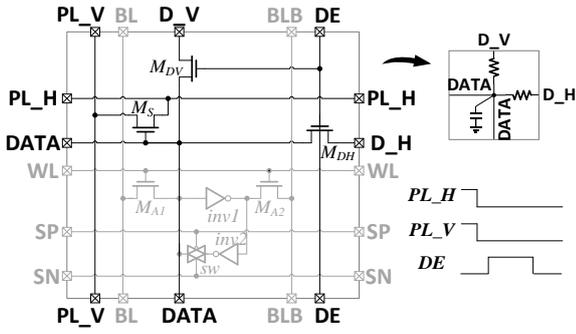

Fig. 2. CRAM bit cell works in IMC based image restoration mode for denoising and filling.

Fig. 2 shows the CRAM working in IMC based IR mode for denoising and filling, where MS acts as a capacitor while MDV and MDH work as resistors controlled by gate signal DE. The bit cell can be modeled by a RC circuit and the array is equivalent to a 2-D RC network, where the charge stored in each cell has nominally identical transmission paths to its neighbors. Initially, the values stored in bit cells are digital signals corresponding to the binary image. With charge diffusion as time goes on, the charge of each cell are redistributed, and the voltages are analog signals (between "0" and "1"). After diffusion is disabled, the first inverter inv1 is used to sense and convert the analog signals to digital signals and then store into the bit cell again for the future processing. The IR effect can be modulated by the signal diffusion enable DE via 1) pulse width 2) pulse amplitude and 3) number of pulses. The CRAM array is surrounded by a dummy mini cell array (as the grey part in Fig 1) to make each cell match well and offer the same diffusion surroundings for charges across the whole array.

Fig. 3 shows the CRAM operating in projection mode for IMC-based RP. The diffusion enable signal DE is deactivated to cut off the charge diffusion path. The switch sw is enabled to latch the data while the world line WL is kept low to disable the memory access. MOSFET MS acts as the projection device for 2-dimensions separately, by which the value stored in SRAM can be readout along rows and columns. To project the data to one projection line PL_H (PL_V), both the orthogonal PLs are pulled down to GND initially and then the PL_H (PL_V) is floating while the PL_V (PL_H) is pulled up to VDD. If the data stored in the cell is "1", the PL_H (PL_V) will be charged and can be sensed out. If the data in the cell is "0", the PL_H (PL_V) will keep floating at zero. For each row and column, a projection detector (PD) is used to configure and detect the projection line. The PD block is composed of a pull-up network (PUN), a pull-down network (PDN), and a sense amplifier (SA). Fig. 3 shows a 320×240 CRAM macro in projection mode together with the PDs of the ith row to detect the horizontal projection line PL_H<i> and of the jth column to detect the vertical projection line PL_V<j>. All the control signals for PUN and PDN are generated from the digital controller block. The reference voltage Vref is an analog voltage programmable by a 4-bit DAC. This voltage is used to compare the projection line voltage and generate the 1-bit detection result.

Fig. 4a illustrates the horizontal projection and detection of an image. The 1st phase of RP using the projections follows iterative and selective search (ISS) algorithm [3] as briefly described next (Fig. 4b). When the PL line voltage is charged over the reference voltage Vref by cells containing "1", this line is detected by the SA and hence the coordinate positions can be found. ISS proceeds by alternatively projecting onto X and Y axes. During a vertical projection, the rows corresponding to object locations found in the previous iteration are enabled (entire axis enabled in 1st projection). The algorithm terminates when the number of objects found in successive iterations are equal. The 2nd phase of RP consolidation is done in the digital controller following Fig. 4c. For each new object detected, it is filtered as noise if its size is smaller than a threshold (SIZE_MIN). Next, it is evaluated for merging by checking if its gaps with previous objects in both dimensions are less than predefined thresholds (SLOT) or not. This helps overcome image fragmentations issue. All the thresholds are programmable to adapt to various application scenarios.

## III. MEASUREMENTS RESULTS

The chip is fabricated in a 65 nm CMOS process. The execution times for different number of objects (N) are measured at 1.2V and 10MHz (Fig. 5a). The execution time depends not only on the number of objects but also on the positions of each object. The minimal execution time can be estimated by 8N+8 for IMC-based RP and 10N+12 including controller cycles. This work demonstrates IMC based global parallel diffusion and column/row-wise projection to achieve a maximal energy efficiency of 1220 TOPS/W for image restoration and 915 TOPS/W when combined with region proposal at 0.8V and 30MHz (Fig. 5b). The F-1 score is evaluated using 70k frames for different IoU values (Fig. 5c) showing superior performance than IMC alone [3] due to the post-processing by the RP update controller. Also, the high accuracy is robust across multiple settings of resistor and diffusion times (Fig. 5c). Variability at bit-cell level is characterized by measuring the diffusion speed at the center and corner locations of 3 chips via a 4x4 blob of "1" pixels written in an image full of "0" s (Fig. 2). The diffusion speed at the center is slightly faster than the corner because charge at the center has larger area to diffuse. Fig.6 compares our prototype IMC-based approach with the state of the art showing superior energy efficiency.

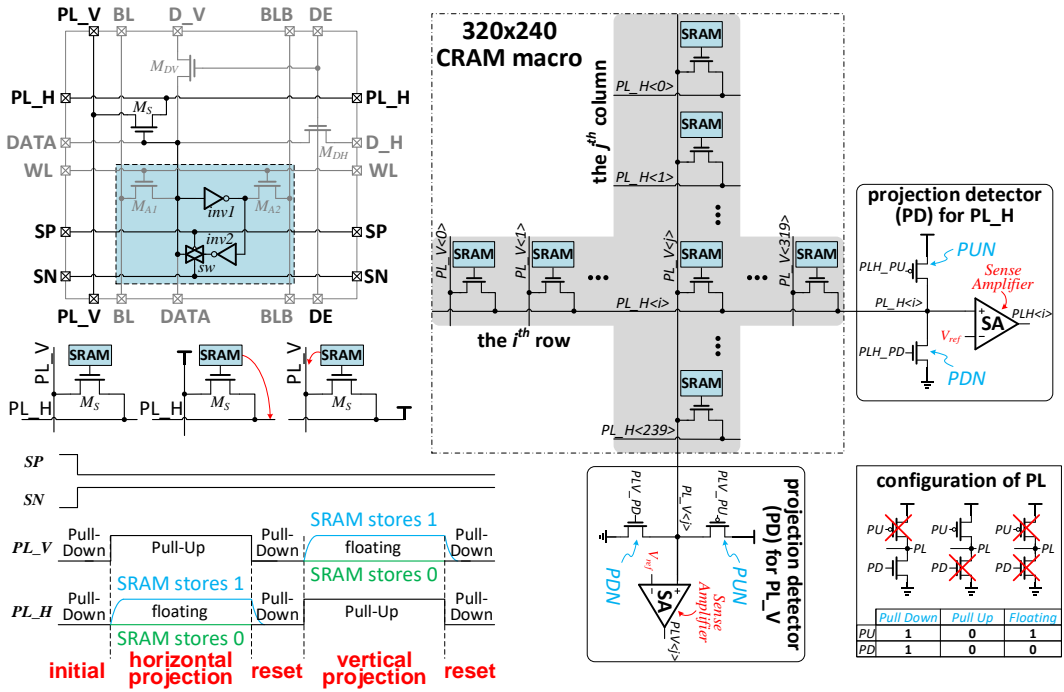

Fig. 3. CRAM operates in projection mode for IMC-based RP and it can be modeled as an SRAM controlled MOS. The bit cell value can be projected to the projection line PL_H (PL_V) by pulling up another projection line PL_V (PL_H) to VDD.

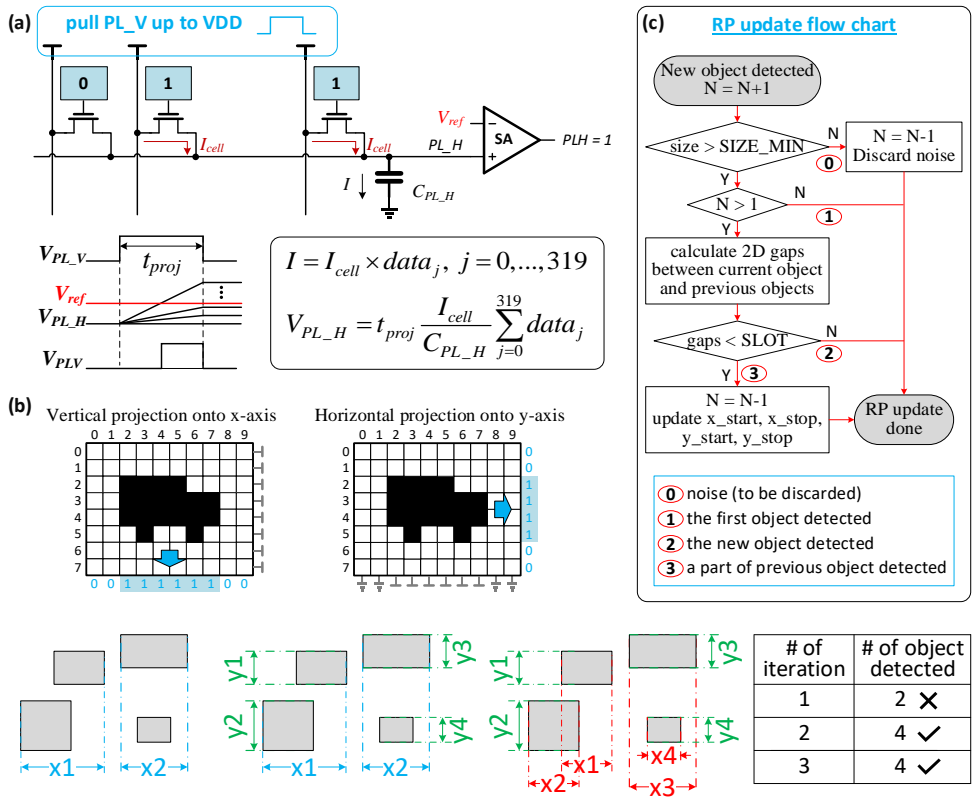

Fig. 4. (a) The projection and detection of one row in IMC-based RP. (b) Example of ISS algorithm. (c) RP update algorithm.

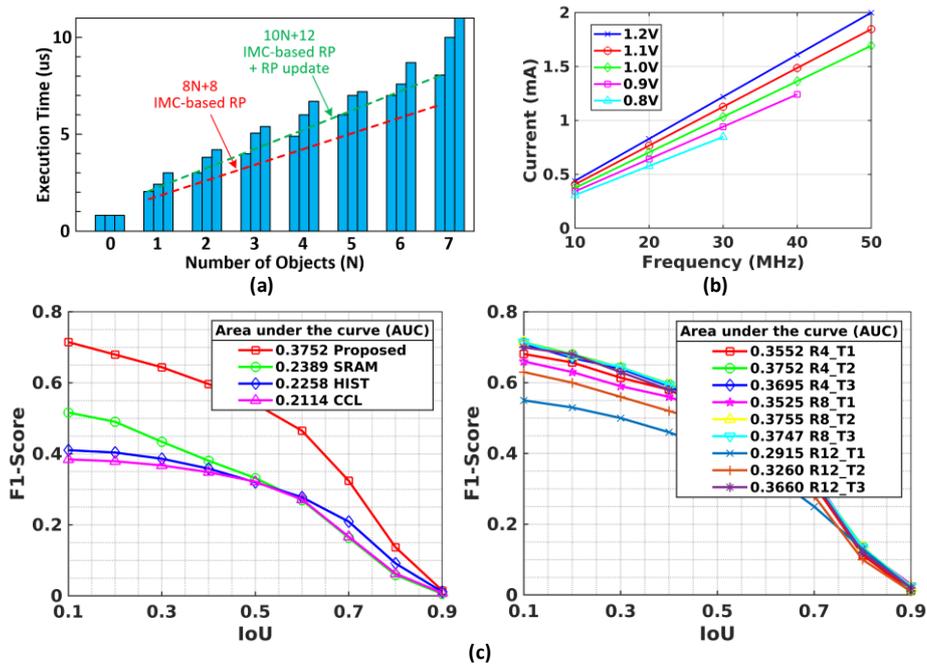

Fig. 5(a) Execution time for different number of objects at 10MHz. (b) Current consumption at different frequencies and supply voltages. (c) Weighted F1-Score comparison with prior works and variation under various resistance and diffusion time settings.

**Comparison with prior In-memory computing works**

| | This work | JSSC'21 [2] | ASSCC'21 [3] | SOVC'20 [4] | JSSC'18 [5] |
|---|---|---|---|---|---|
| Thechnology | 65nm | 65nm | 65nm | 10nm | 65nm |
| Algorithm | Image denoising, Region proposal | Image denoising | Region proposal | Event generation, Region proposal | Versatile DNN |
| Operation Mode | Charge distribution, Current summation | Current summation | Current summation | Digital | Digital XNOR |
| Cell Type | 11T CRAM | 9T SRAM | 9T SRAM | Twin-8T SRAM | 6T SRAM |
| Memory Capacity (kB) | 10.4[a] | 9.375 | 9.375 | 320 | 102 |
| Chip Area (mm2) | 0.62 | 0.55 | 1 | 2.6 | 3.9 |
| Supply (V) | 0.7 – 1.2 | 0.7 – 1.2 | 0.75 – 1.2 | 0.65 – 0.8 | 0.55 - 1 |
| Throughput (GOPS) | 9600 (Denoising) 976(Denoising+RP) | 134.4 | 409 | -[b] | 1380 |
| Energy Eff. (TOPS/W) | 1220 (Denoising) 915(Denoising+RP) | 51.3 | 389 | - | 6.3 |

[a] Including dummy cell ring.
[b] 70fps@1Meps.

Fig. 6. Comparison of the proposed IMC-based IR and RP with other works.